\documentstyle[12pt]{article}

\begin{document}

\title{Quantum Boltzmann statistics in interacting systems}
\author{Luigi Accardi, Sergei Kozyrev}
\maketitle

\centerline{\it Centro Vito Volterra, Universita di Roma Tor
Vergata}

\begin{abstract}
Collective operators for generic quantum system with discrete
spectrum are investigated. These operators, considered as
operators in the entangled Fock space (space generated by action
of collective creations on the vacuum) satisfy a particular kind
of Quantum Boltzmann (or free) commutational relations.
\end{abstract}

\section{Introduction}

In the present paper we investigate the statistics of the
interacting (or entangled) operators in the stochastic limit of
quantum theory \cite{book}. We investigate the model of quantum
system interacting with reservoir (quantum field). The
corresponding Hamiltonian is a combination of interacting, or
entangled operators. Interacting operator is a product of operator
of system and operator of reservoir, for example the product of
annihilation of reservoir (bosonic quantum field) and a system
operator
$$
D^* a(k)
$$
In the stochastic limit the quantum field becomes a quantum white
noise. We will show that for the stochastic limit of discrete
quantum system interacting with quantum field (for details see
\cite{generic}) the statistics of entangled (or interacting)
operators is the particular variant of Quantum Boltzmann, or free
statistics. Analogous behavior (arising of free statistics) was
already found for particular system Hamiltonian with continuous
spectrum \cite{AcLuVo}, \cite{QED}, \cite{Bogconf}, \cite{largeN}.

In the present paper we will investigate generic discrete quantum
system interacting with quantum field.

\bigskip

\noindent {\bf Definition}.\quad{\sl A quantum system with
Hamiltonian $H_S$ acting in the Hilbert space ${\cal H}_S$ will be
called {\it generic\/}, if:
\begin{description}
\item{i)}
The spectrum $\hbox{ Spec }H_S$ of the Hamiltonian is non
degenerate.
\item{ii)}
For any positive Bohr frequency $\omega>0$ there exists a unique
pair of energy levels
$\varepsilon_{1_\omega},\varepsilon_{2_\omega}\in \hbox{ Spec
}H_S$ such that:
$$
H_S|1_\omega\rangle=\varepsilon_{1_\omega}|1_\omega\rangle,\quad
H_S|2_\omega\rangle=\varepsilon_{2_\omega}|2_\omega\rangle
$$
$$
\omega=\varepsilon_{2_\omega}-\varepsilon_{1_\omega}>0
$$
\end{description}
}

\bigskip

The term {\it generic} means that {\it the eigenvalues of $H_S$
are irregularly displaced}. For example, the spectrum of the
1--dimensional harmonic oscillator satisfies (i) but not (ii).
Thus it is not generic in the above sense.

In the stochastic limit, cf. \cite{book}, every positive Bohr
frequency (a difference of eigenvalues of a system Hamiltonian)
gives rise to a quantum white noise.

In the work \cite{generic} the authors showed that, in the
stochastic limit, the dynamics of a (generic) quantum system is
described by a stochastic Schr\"odinger equation which contains
the following combination of quantum noises and system observables
(that we call interacting, or entangled, or collective operator):
\begin{equation}\label{c_jomega}
c_{\omega}(t,k):=|2_{\omega}\rangle\langle 1_{\omega}| \otimes
b_{\omega}(t,k)
\end{equation}
where the quantum white noise, cf. \cite{book}, satisfy the
following (bosonic) relation
\begin{equation}\label{2_5}
[b_{\omega}(t,k),b_{\omega'}(t',k')]=2\pi\delta_{\omega\omega'}\delta(t-t')
\delta(\omega(k)-\omega)\delta(k-k')
\end{equation}
Actually even before the limit the evolution equation in the
interaction picture contains terms of similar kind
$$
|2_{\omega}\rangle\langle 1_{\omega}| \otimes
a(k)e^{-it(\omega(k)-\omega)}
$$
This suggests that Skeide's analysis of the stochastic limit in
terms of Hilbert modules \cite{Ske98} can be extended to the
present case.

In the present paper we show that the operators (\ref{c_jomega})
satisfy a variant of the Quantum Boltzmann (also called free, or
infinite) commutation relations:
\begin{equation}\label{1.29a}
a_i a^*_j=\delta_{ij}
\end{equation}
The {\it Quantum Boltzmann\/} (or {\it free\/}) algebra is
generated by the $a_j$, $a^*_k$, called creation and annihilation
operators, with the relations (\ref{1.29a}). No other relations
are assumed and different creators $a^*_i$, $a^*_j$ do not
commute. Therefore vectors in the Fock space of the type
$$
a^*_i a^*_j |0\rangle  \ne  a^*_j a^*_i  |0\rangle
$$
are distinguishable and that justifies the name {\it Quantum
Boltzmann relations\/} cf. for example \cite{Cuntz},
\cite{Speicher}, \cite{Voiculescu}. Generalizations of such
relations were found in the large $N$ limit of quantum
chromodynamics \cite{largeN}, in models of particles interacting
with a quantum field (which include quantum electrodynamics and
the polaron model) \cite{AcLuVo}, \cite{QED}.

\section{Quantum Boltzmann statistics for entangled operators}

Given a generic quantum system, and a Bohr frequency $\omega$,
denote $|1_{\omega}\rangle$ and $|2_{\omega}\rangle$ the two
eigenstates corresponding to the two energy levels,
$\varepsilon_{1_\omega}, \varepsilon_{2_\omega}$, so that
\begin{equation}\label{1.29b}
H_S|1_{\omega}\rangle=\varepsilon_{1_\omega}|1_{\omega}\rangle\quad
,\quad
H_S|2_{\omega}\rangle=\varepsilon_{2_\omega}|2_{\omega}\rangle\quad
,\quad \omega=\varepsilon_{2_\omega}-\varepsilon_{1_\omega}>0
\end{equation}
With these notations we see that the restriction of $H_S$ on the
space generated by $|1_\omega\rangle$ and $|2_\omega\rangle$ is
$$\pmatrix{
\varepsilon_2&0\cr 0&\varepsilon_1\cr}
$$

We define for each positive Bohr frequency $\omega$ the Pauli
matrix that flips the spin down by
\begin{equation}\label{1.29d}
|1_{\omega}\rangle\langle
2_{\omega}|=:\sigma^{-}_{\omega}=\pmatrix{0&0\cr 1&0\cr}
\end{equation}
Using this we define the interacting (or entangled, or collective)
operator
\begin{equation}\label{1.29f}
c_{\omega}(t,k):=|2_{\omega}\rangle\langle 1_{\omega}|\otimes
b_{\omega}(t,k)= \sigma^+_{\omega} \otimes b_{\omega}(t,k)
\end{equation}

\medskip

\noindent {\bf Remark}.\quad The operator (\ref{1.29f}) defined
for non--positive Bohr frequency $\omega\le 0$ is equal to zero,
since the quantum white noise $b_{\omega}(t,k)$ for $\omega\le 0$
equals to zero.

\medskip

One can get the following relation on the collective operator,
cf. \cite{Ske99}
\begin{equation}\label{ske99}
c^2_{\omega}(t,k)=0
\end{equation}
In the present paper we will investigate the case of many Bohr
frequencies and get for collective operators the Quantum
Boltzmann relations.

\medskip

\noindent {\bf Remark}.\quad
In the paper \cite{Ske99}
operators of the form (\ref{1.29f}) were considered for the case
of a single Bohr frequency and their vacuum statistics was shown to
satisfy Boolean independence in the sense of von Waldenfels.
The present paper extends this result to the case of multiple Bohr
frequencies and establishes a connection between Boolean and
Boltzmannian (or free) independence. It is interesting to notice that
these types of independences arise naturally, i.e. from physically
meaningful objects, in a purely bosonic context.

\medskip

The dynamics of a system in the stochastic limit is described by
the white noise (or master) Hamiltonian which in the considered
case takes the form
\begin{equation}\label{1.29g}
h(t)=\sum_{\omega\in F}\int dk
\overline{g_\omega}(k)\sigma^+_\omega b_\omega(t,k)+\hbox{ h.c.}
\end{equation}
in other words: if $H_S$ is generic, any generalized dipole
interaction Hamiltonian of $S$ with a boson field is equivalent,
in the stochastic limit, to a (possibly infinite) sum of
independent $2$--level--systems. The summation in (\ref{1.29g})
runs over the set of all Bohr frequencies. The simplest case
corresponds to a single $2$--level--system (one Bohr frequency),
or the spin--boson model, cf. \cite{spb}.

Notice that the operators $\sigma^\pm_\omega$, $b^\pm_\omega$
enter in the master Hamiltonian  only through the combinations
\begin{equation}\label{1.29j}
\sigma^+_\omega\otimes b_\omega(t,g)\ ;\quad\sigma^-_\omega\otimes
b^*_\omega(t,g)
\end{equation}
and therefore the basic dynamical quantities like the propagator
$U_t$, the wave operators $\Omega_\pm$, the scattering operator
$S$, will depend only on these combinations. This suggests to
consider algebra generated by the {\it entangled operators\/}
$$c_\omega(t,k):=\sigma^+_\omega\otimes b_\omega(t,k)\ ;\quad
c^*_\omega(t,k):=\sigma^-_\omega\otimes b^*_\omega(t,k)$$ then all
the calculations involving only matrix elements of the iterated
series expansion of $U_t$, the solution of the white noise
Hamiltonian equation, cf. \cite{book}
$$
\partial_tU_t=-i\left(\sum_{\omega\in F}\int dk \overline{g_\omega}(k)
\sigma^+_\omega b_\omega(t,k)+ \hbox{h.c.}\right)U_t
$$
with initial condition $U_0=1$, can be done entirely within this
algebra.

A representation of the {\it entangled algebra\/} can be obtained
within the {\it Fock module\/} ${\cal F}_{\hbox{ent}}$ which is
the submodule of
$${\cal B}({\cal H}_S)\otimes{\cal F}_{\hbox{mast}}$$
where ${\cal F}_{\hbox{mast}}$ is the Fock space of the master
field, ${\cal B}({\cal H}_S)$ is the algebra of bounded operators
in the system Hilbert space ${\cal H}_S$. The Fock module ${\cal
F}_{\hbox{ent}}$ is a linear span of the {\it entangled number
vectors\/}:
$$\prod_n c^*_{\omega_n}(t_n,k_n)|0\rangle$$
where $|0\rangle$ is the Fock vacuum of the master field.

\bigskip

\noindent {\bf Theorem}.\quad {\sl The operators
$c_{\omega}(t,k)$, $c^*_{\omega}(t,k)$ considered as operators on
the Fock module ${\cal F}_{\hbox{ent}}$ of entangled number
vectors satisfy the relations module
\begin{equation}\label{ccdag}
c_{\omega}(t,k) c^*_{\omega'}(t',k')= 2\pi \delta_{\omega\omega'}
\delta(t-t')\delta(k-k')\delta(\omega(k)-\omega)
\sigma^+_{\omega}\sigma^-_{\omega}
\end{equation}
The operator
$\sigma^+_{\omega}\sigma^-_{\omega}=|2_{\omega}\rangle\langle
2_{\omega}|$ in (\ref{ccdag}) is equal to the rank one orthogonal
projector onto ${\bf C}|2_\omega\rangle$. }

\bigskip

\noindent{\bf Remark}. \qquad Notice that the operator
$\sigma^+_{\omega}\sigma^-_{\omega}$ in (\ref{ccdag}) conserves
the entangled Fock space ${\cal F}_{\hbox{ent}}$.

\medskip

\noindent{\it Proof}.\qquad
Using the commutation relation (\ref{2_5}) we get
$$
c_{\omega}(t,k) c^*_{\omega'}(t',k')= 2\pi \delta_{\omega\omega'}
\delta(t-t')\delta(k-k')\delta(\omega(k)-\omega) \sigma^+_{\omega}
\sigma^-_{\omega'}+ \sigma^+_{\omega} \sigma^-_{\omega'}
b^*_{\omega'}(t',k')b_{\omega}(t,k)
$$
Let us show that the operator
\begin{equation}\label{vanishingpart}
\sigma^+_{\omega} \sigma^-_{\omega'}
b^*_{\omega'}(t',k')b_{\omega}(t,k)
\end{equation}
is equal to zero on the Fock module ${\cal F}_{\hbox{ent}}$.
To check this we consider the action of (\ref{vanishingpart}) on a
number vector of
\begin{equation}\label{1.31aa}
\prod_{n=1}^{l} c^*_{\omega_{n}}(t_{n},k_{n})|0\rangle
\end{equation}
The action of (\ref{vanishingpart}) on such a vector will be
non--zero only if in the product (\ref{1.31aa}) at least one of
the creators is equal to $c^*_{\omega}(t,k)$. Let us denote $l_0$
the index of the first creator, starting from the left with this
property. We get
$$
\sigma^+_{\omega} \sigma^-_{\omega'}  \prod_{1\le n <l_0}
\sigma^-_{\omega_n} \sigma^-_{\omega} \prod_{l_0< n \le l}
\sigma^-_{\omega_n} \hbox{ combination of }
b^*_{\omega_n}(t_n,k_n) |0\rangle
$$
But for a generic system the combination
$$
\sigma^+_{\omega} \sigma^-_{\omega'}  \prod_{1\le n <l_0}
\sigma^-_{\omega_n} \sigma^-_{\omega} =|2_{\omega}\rangle\langle
1_{\omega}| \sigma^-_{\omega'} \prod_{1\le n <l_0}
\sigma^-_{\omega_n} |1_{\omega}\rangle\langle 2_{\omega}|
$$
equals to zero since
\begin{equation}\label{1Eomega1}
\langle 1_{\omega}| \sigma^-_{\omega'}  \prod_{1\le n <l_0}
\sigma^-_{\omega_n} |1_{\omega}\rangle=0
\end{equation}
To prove this let us note that for generic Hamiltonian every
operator $\sigma^-_{\omega_k}$ acting on eigenvector
$|1_{\omega}\rangle$ of $H_S$ (with the eigenvalue $\varepsilon$)
kills it or maps into the eigenvector $|2_{\omega}\rangle$ with
the eigenvalue $\varepsilon-\omega$. Since arbitrary operator
$\sigma^-_{\omega_k}$ decreases the energy (this is exactly the
place where the stochastic limit procedure is important, since in
the stochastic limit only the terms with positive Bohr frequences
in interaction Hamiltonian survive), and the product of
$\sigma^-_{\omega_k}$ in (\ref{1Eomega1}) is non--empty, the
result of application of the product of $\sigma^-_{\omega_k}$ in
(\ref{1Eomega1}) to $|1_{\omega}\rangle$ can be either zero or
vector orthogonal to $|1_{\omega}\rangle$. Therefore the matrix
element (\ref{1Eomega1}) is equal to zero.

We checked that the second term (\ref{vanishingpart}) vanishes
that finishes the proof of relations (\ref{ccdag}).\medskip

\medskip

\noindent{\bf Remark}.\quad One can find an alternative
representation for the relations (\ref{ccdag}), defined by the
operators
\begin{equation}\label{newreps}
c_{\omega}(t,k)= \sigma^+_{\omega} c(t)\otimes c_{\omega}(k)
\end{equation}
where $c(t)$, $c_{\omega}(k)$ satisfy the relations
$$
c(t)c^*(t')=\delta(t-t')
$$
\begin{equation}\label{modulerelat}
c_{\omega}(k)c_{\omega'}^*(k')=2\pi\delta(\omega(k)-\omega)
\delta_{\omega\omega'}\delta(k-k')
\end{equation}
This gives another representation of (\ref{ccdag}) using the
tensor product of Quantum Boltzmann algebras. An analogous
representation for the QED (quantum electrodynamics) entangled
algebra was proposed in \cite{QED}.  Moreover the operator
$c_{\omega}(k)$ in (\ref{modulerelat}) can be formally reproduced
by the formula
$$
c_{\omega}(k)=  c_{\omega}\otimes\sqrt{2\pi\delta(\omega(k)-\omega)}c(k)
$$
where $c_{\omega}$ and $c(k)$ are Boltzmannian annihilators.

\medskip

\noindent{\bf Remark}. In the paper \cite{Bogconf} the operator
valued Quantum Boltzmann relations arising in the stochastic
limit were called {\it entangled\/}, or {\it interacting\/}
commutation relations.

\medskip

\noindent{\bf Remark}. In \cite{Bogconf} a second quantized
version of the module Quantum Boltzmann relations have been shown
to be universal in any interaction with conservation of momentum.

\medskip

The results of the present paper show that such relations arise in
every generic open quantum system. This rises the problem of
extending the above construction to non--generic systems thus
obtaining a classification of all the possible commutation
relations arising in the stochastic limit of quantum mechanics.

\medskip

\noindent{\bf Remark}. The term {\it entanglement\/} is usually referred
to states and expresses the impossibility of representing a state vector
of an open system as a tensor product, i.e. in the form
$$
\psi_{system}\otimes\psi_{reservoir}
$$
Such states are unstable under the evolution of interacting systems which
creates linear combinations of them, i.e. states of the form:
$$
\sum_n \psi^{(n)}_{system}\otimes\psi^{(n)}_{reservoir}
$$
(with a number of terms strictly greater than $1$) these are the
entangled states. In the stochastic limit, the evolution of
generic interacting systems (that creates entanglement) is
described by Quantum Boltzmann relations of the type
(\ref{ccdag}) for collective operators (system plus reservoir).
Therefore we might say that entangled states present the
Schr\"odinger picture of entanglement while the entangled
relations correspond to the Heisenberg picture of entanglement.

\bigskip

\centerline{\bf Acknowledgements}

The authors are grateful to I.V.Volovich and M.Skeide for
discussions. Sergei Kozyrev is grateful to Centro Vito Volterra
and Luigi Accardi for kind hospitality. This work was partially
supported by INTAS 9900545 grant. Sergei Kozyrev was partially
supported by RFFI 990100866 grant.

\end{document}